\def\ni{\noindent}
\def\Im{I_{{\rm m}}}

\def\rs{{\rm s}}
\documentstyle[epsf]{mn}
\title[Cusps without Chaos]
{Three dimensional, axisymmetric cusps without chaos}
\author[S.~Sridhar and J.~Touma]
	{S.~Sridhar\thanks{E-mail: sridhar@iucaa.ernet.in} and
	 J.~Touma\thanks{E-mail: touma@harlan.as.utexas.edu}\\
	$^*$~Inter-University Centre for Astronomy and Astrophysics, 
        Ganeshkhind, Pune 411 007, INDIA\\
	$^{\dag}$~University of Texas, McDonald Observatory, 
	RLM 15.308, Austin, Texas, 78712}

\date{IUCAA Preprint 28/97, March 1997}
 
\pagerange{\pageref{firstpage}--\pageref{lastpage}}
 
\begin{document}
\label{firstpage}
\maketitle

\begin{abstract}
We construct three dimensional axisymmetric, cuspy density
distributions, whose potentials are of St\"ackel form in parabolic
coordinates. As in Sridhar and Touma~(1997), a black hole of 
arbitrary mass may be added at the centre, without destroying the St\"ackel 
form of the potentials. The  construction uses a classic method, originally
due to Kuzmin~(1956), which is here extended to parabolic coordinates.
The models are highly oblate, and the cusps are ``weak'', with the 
density, $\rho\propto 1/r^k$, where $0<k<1\,$.
\end{abstract}

\section{INTRODUCTION}
Kuzmin(1956) showed that the density at a general point of an oblate,
axisymmetric mass model, whose gravitational potential is of St\"ackel
form in elliptic coordinates, is simply related to the density on the
axis of symmetry. Later, Kuzmin~(1973) generalized the theorem to the
fully triaxial case. A corollary of Kuzmin's
theorem is that the density so constructed is automatically
non-negative everywhere as long as it is non-negative on the short
axis. 
The interest in St\"ackel models arises from the
fact that the motion of point masses in such potentials is completely
integrable by the Hamilton--Jacobi method; these potentials are also
termed ``separable'', since the dynamics in three dimensions separates
into three independent oscillations (c.f. de~Zeeuw~1985a,b for a clear
discussion of orbits, and Kuzmin's theorem). Constructing distribution
functions for models of galaxies in a steady state (c.f. Binney \&
Tremaine~1987) is facilitated by the existence of three isolating
integrals for three dimensional models.   
Recently, Sridhar and Touma~(1997)---hereafter referred to as
ST---presented a family of cuspy, scale--free mass models of
non-axisymmetric discs whose potentials are of St\"ackel form in
parabolic coordinates. The surface density behaves like $\Sigma\propto
1/r^{\gamma}$, where $0 <\gamma <1$. Here we generalize
Kuzmin~(1956) to parabolic coordinates and naturally extend the
non-axisymmetric discs of ST to three dimensional, axisymmetric, cuspy
density distributions, whose potentials are of St\"ackel form in
parabolic coordinates. Our main result follows in
\S~2. Scale--free cusps are presented in \S~3, and a complete
classification of orbit families is given in \S~4.  A discussion of
the results, in \S~5, completes this paper on mass models. 

\section{KUZMIN'S CONSTRUCTION APPLIED TO PARABOLIC COORDINATES}

For axisymmetric configurations, the mass density $\rho$, and the
gravitational potential $U$, that it generates are both functions only
of $R=\sqrt{x^2 + y^2}$ and $z$, the cylindrical radius and the
coordinate along the axis of symmetry, respectively. Since the
potential is independent of the azimuthal angle ($\phi$), the component of
the angular momentum parallel to the $z$--axis, $L_z$, is
conserved. Hence the dynamics may be described as two--dimensional
motion in a meridional ($\phi=\hbox{constant}$) plane, under the
influence of an effective potential defined by

\begin{equation}
U_{{\rm eff}}=U(R, z) + \frac{L_z^2}{2R^2}.\label{effpot}
\end{equation}
\ni We study cases when the motion in the meridional plane is
separable in parabolic coordinates, 
$(\xi, \eta)$. These may be
defined as the roots for $\tau$ of the
equation

\begin{equation}
\tau^2 -2z\tau -R^2=0.\label{quad}
\end{equation}
\ni Solving this quadratic equation gives

\begin{eqnarray}
\xi &=& z + \sqrt{z^2 + R^2} \nonumber \\
\eta &=& z - \sqrt{z^2+R^2},\label{partocyl}
\end{eqnarray}
where we have chosen $\eta\leq 0\leq\xi$. The inverse relations are

\begin{equation}
R^2 = -\xi \eta,\quad\quad z=\frac{1}{2}(\xi +\eta),\quad\quad
r = \frac{\xi-\eta}{2},\label{cyltopar}
\end{equation}
\ni where $r=\sqrt{R^2 + z^2}$ is the spherical radius.
Surfaces of constant $\xi$ and $\eta$ are paraboloids of revolution which
cut the $z$-axis at $\xi/2$ and  $\eta/2$, and the $x$-$y$ plane at $R=\xi$ and 
$R=\eta$ respectively. These surfaces are orthogonal to the meridional plane
and mutually orthogonal as well; thus $(\xi, \eta, \phi)$ form an orthogonal
coordinate system. Positive and negative ranges
of the $z$--axis are covered by $\xi$ and $\eta$ respectively:

\begin{eqnarray}
\xi=0, & z=2\eta, & \mbox{for}\; z \leq 0 \nonumber \\
\eta=0, & z=2\xi, & \mbox{for}\; z \geq 0 .\label{zaxis}
\end{eqnarray}

The most general axisymmetric potential in parabolic
coordinates, for which the Hamilton--Jacobi equation separates,
has the St\"ackel form,  given by
\begin{equation}
U(\xi, \eta) = \frac{F_{+}(\xi)}{\xi - \eta} + \frac{F_{-}(\eta)}{\eta - \xi}.\label{stackel}
\end{equation}
\ni If $U(\xi, \eta)$ is of St\"ackel form, 
so is the effective potential, $U_{{\rm eff}}$, which is defined in
equation~(\ref{effpot}).\footnote{To verify this we only need note
that adding $L_z^2/2\xi$ and $L_z^2/2\eta$ to $F_{+}$ and $F_{-}$
respectively, adds the centrifugal term to $U(\xi, \eta)\,$.}~The
potential due to a point mass is obtained by choosing
$F_{-}-F_{+}=2GM$. It is important to note that, to describe an
interesting case like the Kepler problem, $F_{+}\neq F_{-}$; in
general, we let $F_{+}$ and $F_{-}$ be quite different functions of
their arguments. The density corresponding to the point mass assumes
the form,
\begin{equation}
\rho_{\bullet}\equiv M\delta({\bf r})=
M\left\{\frac{2\delta(\xi)\delta(\eta)}{\pi(\xi -\eta)}
\right\}\,\label{bhdensity}
\end{equation}

\ni Let $\rho$ and $U$ be the density and potential, respectively, of 
a smooth distribution of matter\footnote{We include density cusps
in our ``smooth'' distribution!}. If we now include a point mass at the
origin, 
\begin{eqnarray}
\rho_{tot} &=& \rho +\rho_\bullet \nonumber \\
U_{tot} &=& U -\frac{2GM}{\xi -\eta},\label{total}
\end{eqnarray}

\ni are the total density and potential. These are related to each other
through Poisson's equation, or
equivalently
\begin{equation}
\nabla^2U=4\pi G\rho,\label{poisson}
\end{equation} 

\ni where the contribution of the point mass has been dropped from both sides.
Expressed in parabolic coordinates, Poisson's equation for 
(the axisymmetric) potential and density reads

\begin{equation}
\frac{4}{\xi-\eta} \left [ \frac{\partial}{\partial \xi} \xi
\frac{\partial}{\partial \xi} - \frac{\partial}{\partial \eta} \eta
\frac{\partial}{\partial \eta} \right ] U(\xi, \eta) = 4 \pi G
\rho(\xi, \eta).\label{poissonpar}
\end{equation}

\ni We substitute the expression given in equation~(\ref{stackel})
for $U$, in equation~(\ref{poissonpar}). This  results in the following
expression for the smoothly distributed density of matter:

\begin{equation}
\pi G\rho(\xi,\eta) =
\frac{\xi F^{''}_{+} +\eta F^{''}_{-}}{(\xi-\eta)^2}
 - \frac{\xi+\eta}{(\xi-\eta)^3} (F^{'}_{+} -
F^{'}_{-})\,,\label{density}
\end{equation}

\ni where the primes denote differentiation with respect to the arguments,
$\xi$ or $\eta$, as the case may be. It proves convenient to employ the single
variable $\tau$ in place of the two variables $\xi$ and $\eta$. Then the
density is completely determined by
\begin{eqnarray}
F(\tau)=\left\{ \begin{array}{ll}
		F_{+}(\xi) & \mbox{if $\tau\geq 0$} \\[1ex]
		F_{-}(\eta) & \mbox {if $\tau\leq 0$}
		\end{array}
	\right. ,\label{ftau}
\end{eqnarray}

\ni which is evidently one function of one variable. It is also useful to define
constants $C_\tau$ and $D_\tau$, that depend only on the behaviour of $F$ for 
small $|\tau|$:
\begin{eqnarray}
C_\tau=\left\{ \begin{array}{ll}
		C_+\equiv\lim_{\eta\rightarrow 0}\eta F_{-}^{'} &
			\mbox{if $\tau\geq 0$} \\[1em]
		C_- \equiv\lim_{\xi\rightarrow 0}\xi F_{+}^{'} &
			\mbox{if $\tau\leq 0$}
		\end{array}
	\right.,\label{constantc}
\end{eqnarray}
\begin{eqnarray}
D_\tau=\left\{ \begin{array}{ll}
		D_+\equiv\lim_{\eta\rightarrow 0}(\eta F_{-}^{'})^{'} &
			\mbox{if $\tau\geq 0$} \\[1em]
		D_- \equiv\lim_{\xi\rightarrow 0}(\xi F_{+}^{'})^{'} &
			\mbox{if $\tau\leq 0$}
		\end{array}
	\right.,\label{constantd}
\end{eqnarray}

We are now ready to apply  Kuzmin~(1956)'s method. Let us assume that the
density profile on the $z$--axis, $\psi(\tau)$, is specified:
\begin{eqnarray}
\psi(\tau)=\left\{ \begin{array}{ll}
		\psi_{+}(\xi)\equiv\rho(\xi, 0) & \mbox{if $\tau\geq 0$} \\[1ex]
		\psi_{-}(\eta)\equiv\rho(0, \eta) & \mbox {if $\tau\leq 0$}
		\end{array}
	\right. ,\label{psitau}
\end{eqnarray}

\ni Then we can use equation~(\ref{density}) to first
determine $F(\tau)$, and hence $\rho(\xi, \eta)$. Restricting
equation~(\ref{density}) to the $z$--axis, results in the following 
(ordinary) differential equation for $F(\tau)\,$:
\begin{equation}
F^{''} - \frac{F^{'}}{\tau} = \pi G \tau
\psi(\tau)-\frac{C_\tau}{\tau^2}-\frac{D_\tau}{\tau}\,,\label{diffeqn}
\end{equation}

\ni where primes denote differentiation with respect to
$\tau$.  We define $\Psi(\tau) =
\int_{0}^{\tau} \psi(s)\,ds\:$, and integrate equation~(\ref{diffeqn})
twice, to obtain the general solution,

\begin{equation}
F(\tau) = \pi G \int_{0}^{\tau} s \Psi(s) ds
+A_\tau + B_{\tau}\frac{{\tau}^{2}}{2} +\frac{C_\tau}{2}\ln{|\tau|}
+D_\tau\tau\,,\label{solution}
\end{equation}

\ni where $A_\tau$ and $B_\tau$ are constants of integration.
We recall, from the discussion following equation~(\ref{stackel}),
that $A_{\eta} -A_{\xi}$ will merely add to $M$;
to describe a smooth mass distribution, we may,
without loss of generality, set  $A_{\eta}=A_{\xi}=0\,$.
Substituting this solution for $F(\tau)$ in equation~(\ref{density})
provides us with a general expression for the St\"ackel 
density (whose $z$--axis profile is $\psi(\tau)$) in parabolic coordinates:
\begin{eqnarray}
\rho(\xi, \eta) &=& \frac{\xi^2\psi_{+}(\xi) + \eta^2\psi_{-}(\eta)}{(\xi -\eta)^2} \nonumber \\
&-&\frac{2\xi\eta}{(\xi -\eta)^3}\left[\Psi_{+}(\xi) -\Psi_{-}(\eta)\right] \nonumber \\
&+& \rho_{{\rm bcd}}\,.\label{eqndensity}
\end{eqnarray}

\ni We recall that the St\"ackel nature of the potential---hence 
separability of the Hamilton--Jacobi equation---remains unaffected if a
point mass (see eqns.~[\ref{bhdensity}] and [\ref{total}]) with arbitrary 
$M$ is added at the origin. In equation~(\ref{eqndensity}), $\rho_{{\rm bcd}}$ 
is the contribution from the $B_\tau$, $C_\tau$, and $D_\tau$ terms in
equation~(\ref{solution}). We show in the Appendix that this
contribution is generally unphysical, so we ignore
$\rho_{{\rm bcd}}$ in the rest of the paper. It
is evident that the remaining expression for $\rho(\xi,
\eta)$ is non negative.
 
\section{Scale--Free Cusps}
A $z$-axis density profile  of
the form $\psi(\tau)={\rho}_0\left(r_0/|\tau|\right)^{k}$,
can give rise to a finite, non--negative St\"ackel density only if
$0<k<1$; this is so because we require cuspy
densities to have $k >0$, and only when $k <1$ is 
\begin{equation}
\Psi(\tau)\equiv\int_{0}^{\tau} \psi(s)\,ds= \frac{\rho_0 r_0^k}{(1-k)}
\Theta(\tau) |\tau|^{1-k}\,,\label{bigpsi}
\end{equation}
\ni finite. Here, $\Theta(\tau)$ is $\pm 1$ accordingly as $\tau$ is positive,
or negative. $F(\tau)$ is readily computed using equation~(\ref{solution}).
Scale--free cusps, with a central black hole are described by 
\begin{eqnarray}
F_{\rs +}(\xi) &=& 2K\xi^{3-k} - GM\nonumber \\[1ex]
F_{\rs -}(\eta) &=& -2K|\eta|^{3-k} + GM\,\label{sfcusps}
\end{eqnarray}
\ni where $K=\pi G\rho_0 r_0^k/2(1-k)(2-k)$. The potential--density pairs
have simple expressions in parabolic coordinates,
\begin{eqnarray}
U_\rs(\xi, \eta) &=& \frac{2K[\xi^{3-k} +
|\eta|^{3-k}] - 2GM}{\xi-\eta}\nonumber\\[1em]
\rho_\rs(\xi, \eta) &=& \rho_0 r_0^k\,\frac{\xi^{2-k} + {|\eta|}^{2-k}}{(\xi -\eta)^2} \nonumber \\
& &-\frac{\rho_0 r_0^k}{(1-k)}\frac{2\xi\eta}{(\xi -\eta)^3}\left[{\xi}^{1-k} + {|\eta|}^{1-k}\right] \nonumber \\
& & + M\delta({\bf r})\,, \label{sfpotden}
\end{eqnarray}
\ni or equally well in spherical polar coordinates:
\begin{eqnarray}
U_\rs(r, \theta) &=& Kr^{2-k}[\,(1+\cos\theta)^{3-k} + (1-\cos\theta)^{3-k}\,]
-\frac{GM}{r}\nonumber\\[1em]
\rho_\rs(r, \theta) &=& \rho_0\left(\frac{r_0}{r}\right)^k[\,(2-k-\cos\theta)(1+\cos\theta)^{2-k} \nonumber \\
& & + (2-k + \cos\theta)(1-\cos\theta)^{2-k}\,] \nonumber \\
& & + M\delta({\bf r})\,.\label{sfpolar}
\end{eqnarray}
\ni Meridional sections of the 
isocontours of the volume density and potential for $k$ equal to $0.1$
and $0.5$ are shown in Figure~1. For definiteness, the section may be
taken to be the $x$-$z$ plane.  The density (Figures 1a and 1c) is
stacked on highly oblate figures of revolution, that display a dip
near the $z$ axis. In all figures, the black hole is shown as a dot at
the centre. In Figures 1b and 1d, the contribution of the black hole
to the potentials is suppressed.  As in the corresponding figures of
ST, the potential isocontours have $k$--independent axis ratio of 2.
\begin{figure}
\epsfxsize=3.3truein\epsfbox[76 200 564 690]{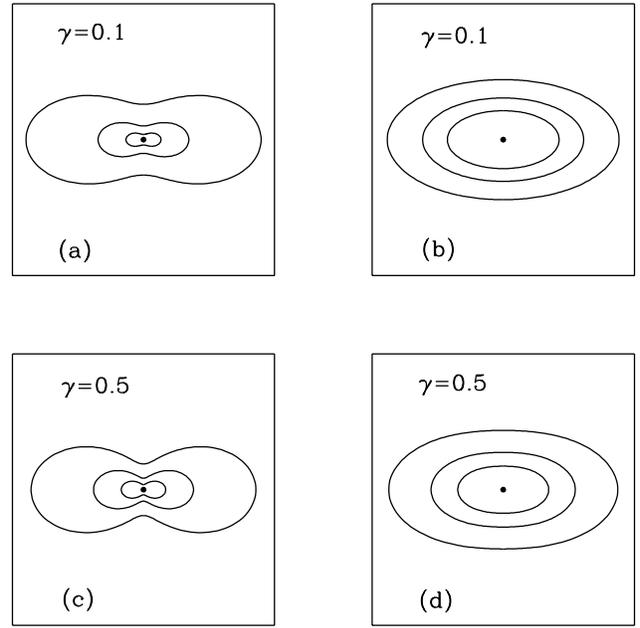}
\caption{Isocontours of Density and Potential:
Figures (a) and (c) show isocontours of the surface density for
two different values of $\gamma$. The corresponding potentials
are displayed in the panels on the right in (b) and (d).
Successive isocontours of the density have ratios of $1.1$ and $1.5$ in (a)
and (c) respectively. Successive potential isocontours
in (b) and (d) have ratios of 2. The location of the
central black hole is shown as a solid dot in all four figures,
although the contribution to the potentials is not included
in (b) and (d).}
\end{figure}

The motion of point masses in  
all axisymmetric potential respects two classical, isolating integrals of
motion: $(E, L_z)$, the energy and the angular momentum about the $z$--axis.
St\"ackel potentials are special in that the Hamilton--Jacobi equation
separates in special, orthogonal coordinates, giving rise to a third
isolating integral $I_3$ (c.f. Landau \& Lifshitz~1976). For motion in the meridional plane, $L_z$ may be treated as a given constant. We give below, 
without derivation, expressions for $E$ and $I_3$:
\begin{equation}
E = \left(\frac{2\xi}{\xi -\eta}\right)p_{\xi}^2 +
\left(\frac{2\eta}{\eta -\xi}\right)p_{\eta}^2 +U_{{\rm eff}}(\xi, \eta)\,,
\label{energy}
\end{equation}
\begin{eqnarray}
I_3 &=&2\xi p_{\xi}^2 + F_{+}(\xi) +
\frac{L_z^2}{2\xi} - E\xi\nonumber\\[1em] 
&=&2\eta p_{\eta}^2 + F_{-}(\eta) +
\frac{L_z^2}{2\eta} - E\eta 
\,.\label{third}
\end{eqnarray}

\section{Orbits}
St\"ackel potentials being integrable, orbits are confined to 3--tori
in the 6--dimensional phase space. The three isolating integrals, $E$, $L_z$ 
and $I_3$, allow us to classify orbit families. When the black hole is 
present, $E$ can take all real values, whereas only positive values are permited
for cusps without black holes\footnote{No fundamental significance is
attached to these positive values of $E$; they merely reflect the fact that our 
cuspy potentials have been chosen to be non negative.}
For  given $E$,  the magnitude of $L_z$
must be less than, or equal to the angular momentum of the circular,
equatorial orbit of the same energy. Having fixed $E$ and $L_z$, 
the third integral, $I_3$, will determine the excursions in $\xi$ and 
$\eta$. From equation~(\ref{third}), the requirement that $p_{\xi}^2$
and $p_{\eta}^2$ be non negative gives
\begin{equation}
I_3\geq g(\xi),\quad\quad\quad -I_3\geq g(|\eta|)\,,\label{range}
\end{equation}
\noindent
where
\begin{equation}
g(s)= 2Ks^{3-k} + \frac{L_z^2}{2s} -Es -GM\,;\quad\quad s\geq 0\,.
\label{funcg}
\end{equation}
\ni Thus, the range of $I_3$ is determined by the minimum of the 
function $g(s)$, which is necessarily non positive. If this 
minimum value is denoted by $-\Im(E, L_z^2)$, we obtain the condition
$-\Im\leq I_3 \leq \Im\,$. We note that the dynamics is truly scale--free only
in the absence of the black hole. 

Given $(E, L_z, I_3)$, an orbit belonging to this 3--torus can explore 
a region of real space, whose boundaries are determined by the intersections
of coordinate surfaces. $x$-$z$ cuts of these regions, for representative
orbit families, are shown shaded in Figures 2a--2d.  For definiteness, we have
chosen a scale--free potential of the form given in equation~(\ref{sfpotden}).
When $I_3$ flips sign, the allowed region for the orbit is reflected about the 
$x$-$z$ plane; hence we will consider only $I_3\geq  0\,$.  

\ni 1. The $L_z\neq 0$ orbits are all {\em loops} about the $z$--axis. These 
avoid the $z$-axis, and generally fill a region of space that is
rotationally symmetric about the $z$-axis; Figure~2a displays the intersection
of such a region, for a generic loop, with the $x$-$z$ plane. When $I_3=0$,
the region is obviously symmetric about the $x$-$y$ plane, as is evident
from Figure~2b. The shaded region now is filled with resonant orbits; the two
frequencies, for motion in the meridional plane, are equal.

\ni 2. Polar orbits, obtained when $L_z=0$, display greater variety.
Motion is restricted to a plane which, for convenience, we think of as
the $x$-$z$ plane, and the orbits can cross the $z$-axis.
In this plane, the orbits are similar to
those that live in the phase space of the non-axisymmetric
disc potentials considered by ST. To keep the
presentation self-contained, we go over the classification here:

\ni (i) $E < 0\:$: $g(s)$ increases monotonically with $s$,
reaching a minimum value of $-GM$ for $s=0$; thus 
$-GM < I_3 < GM$. For a fixed (say, positive) value
of $I_3$, the motion is bounded by the coordinate curves
\begin{eqnarray}
0\leq &\xi &\leq \xi_m\,,\nonumber \\
0\leq &|\eta|&\leq\eta_m\,\label{lensrange}
\end{eqnarray}
\ni where $\xi_m(E, I_3) > \eta_m(E, I_3) >0$ are the two roots
of $g(s)=\pm I_3$. The orbits, named {\em lenses} by ST, fill a lenticular region bounded by the 
parabolas $\xi=\xi_m$ and $\eta=-\eta_m$. As the shaded region in
Figure~2c indicates, the lens orbits visit the origin. 
When $|I_3|=GM$, its maximum value, the shaded region collapses to 
an interval of the $z$--axis, $0\leq z\leq \xi_m/2$. 
For negative values of $I_3$, the roles of $\xi$ and $\eta$ are interchanged.

\ni (ii) $E > 0\:$: $g(s)$ is not monotonic
in $s$; a minimum, $-\Im < 0$, is attained at
$s=[E/2K(3-k)]^{1/(2-k)}$.  Lenses occur for $|I_3| < GM$, but a new
type of orbits appears when $GM < |I_3| \leq \Im$: the orbits are
centrophobic bananas (they cross the $z$-axis, but avoid the origin),
and the allowed region is shown in Figure~2d. They are parented by a
resonant {\em banana} orbit which is obtained when $I_3=\Im$.  When
$M_\bullet=0$, $E$ is always positive; the lens orbits disappear,
leaving the bananas as the only generic family of orbits.

\begin{figure}
\epsfxsize=3.3truein\epsfbox[76 200 564 690]{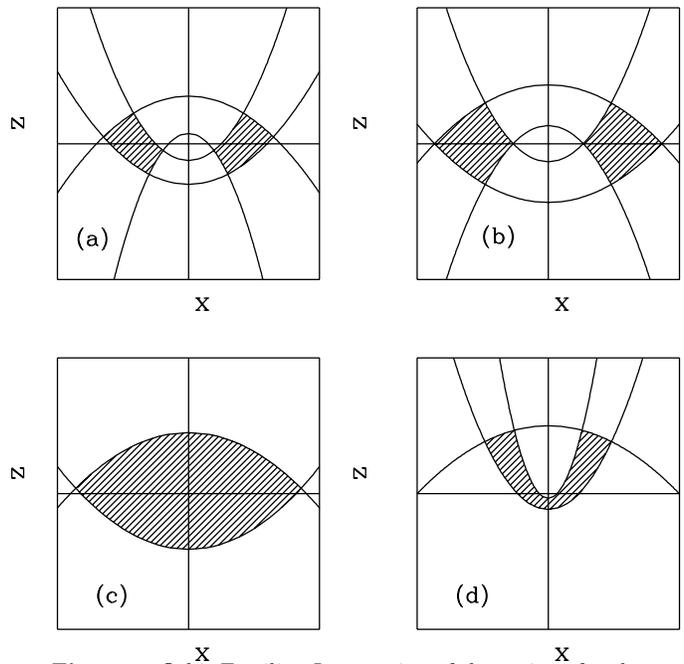}
\caption{Orbit Families: Intersection of the region of real space filled
by orbits on a 3--torus (in phase space) with the $x$-$z$ plane is
shown as the shaded area in all the figures. (a) and (b) correspond to
loop orbits which have $L_z\neq 0$. (a) is for a ``generic'' set of
loops, each of which densely fills the shaded region.  (b) corresponds
to loops with $I_3=0\,$; these resonant, simple, closed (1:1)
orbits. Both (c) and (d) are polar orbits with $L_z=0$. These live
entirely on one meridional plane, here taken to be the $x$-$z$
plane. (c) and (d) are filled by centrophilic lenses, and centrophobic
bananas, quite similar to those discussed in ST.}
\end{figure}

\section{Discussion}
We have generalized Kuzmin's theorem to parabolic coordinates and
constructed three dimensional, axisymmetric, cuspy mass models whose
potentials are of St\"ackel form. A point mass may be added at the
centre, while leaving the separable (i.e. St\"ackel) nature of the
potentials unimpaired.  We have also presented a set of scale--free
cusps, $\rho\propto 1/r^k$, where $0< k < 1$, corresponds to the weak
cusps seen in the ``core--type'' elliptical galaxies of Lauer
et. al~(1995).
 
Kuzmin's original construction in elliptic coordinates allows the
central axis ratio to be specified as an independent parameter,
whereas the very choice of parabolic coordinates forces the
isopotential contours (when the contribution of the black hole to the
potential is neglected) to be significantly oblate, with axis ratio
equal to 2. There are some other essential differences between models
in parabolic coordinates on the one hand, and those studied earlier in
elliptic/ellipsoidal coordinates. As de~Zeeuw, Peletier \&
Franx~(1986) noted, St\"ackel models in ellipsoidal coordinates do not
give rise to cuspy densities; furthermore, the St\"ackel nature is
completely destroyed if a point mass is included at the centre. Our
models are fortunate to escape these limitations. However, the
axisymmetric models we have been able to construct have densities that
can be no steeper than $1/r$; this not only limits our models to weak
cusps, but does not allow construction of finite mass models. On the
other hand, the St\"ackel models in ellipsoidal coordinates have
densities that fall off as steeply as $1/r^4$ far from the centre.

When the potential of the black hole is neglected, the potential
isocontours of our models are similar, concentric ovals.  They may be
compared with the potentials of the power--law galaxies of
Evans~(1994), which are stratified on similar, concentric spheroids.
However, the dynamics in the power--law models is not integrable
(there is no third integral). And like all scale--free models they
have infinite mass.  Evans' oblate, power--law galaxies also have
isodensity contours that are dimpled near the $z$--axis (see also
Richstone~1980, and Toomre~1982 for discussions of scale--free models
with flat rotation curves; with $k=2$ for these models, the cusps are
very strong indeed).  Evans notes that there is no real evidence that
the density distribution of ellipticals is not dimpled, and further
adds that his strongly dimpled models have projected surface density
isocontours that are somewhat boxy.  Since our models have only weak
cusps, the surface density can be computed only with the specification
of a truncation at large $r$. While it may be tempting to speculate
that our St\"ackel models capture some essential features of the
dynamics in the nuclear regions of giant ellipticals (whose isophotes
are known to be boxy), we must be cautious: truncation at large $r$
will likely spoil the exact separability that our models presently
enjoy. Whether one can effect a truncation that retains much of the
integrable nature is yet to be explored.

While three dimensional, axisymmetric cusps are not expected to be
chaotic (c.f. Richstone~1982), it is nevertheless useful to have
explicit integrable mass models with simple analytic expressions (for
the density and potential) that also include a central black hole. The
loops are the only generic orbit family, a fact that was recognised by
Ollongren~(1962) in his classic study of orbits in galactic
potentials. \footnote{Having determined the boundary of the region (of
the meridional plane) in which the orbit is confined, Ollongren~(1962)
loses interest in parabolic coordinates as being useful for a
description of integrable stellar motion in our Galaxy.} Polar orbits
fall in two categories: the centrophobic bananas and the centrophilic
lenses. The models presented in this paper are the natural extension
of the non-axisymmetric two dimensional cusps that we have recently
presented in ST. Together they exhaust the family of scale--free cusps
that can be constructed in parabolic coordinates. Integrable, triaxial
cusps still pose a challenge.

\section{Acknowledgments}
We thank Dave Syer for comments on the manuscript. JT acknowledges the support
of NASA grant NAGW 1477.
\appendix
\section{}
We describe the contribution to $\rho$ of the $B_{\tau}$, $C_{\tau}$
and $D_{\tau}$ terms in Eq.~(\ref{solution}). Writing
$\rho_{{\rm bcd}}=\rho_{{\rm b}} + \rho{{\rm c}} + \rho{{\rm d}}$, 
the three contributions in order are:

\begin{equation}
\rho_{{\rm b}}(\xi, \eta) = \frac{B_{+} - B_{-}}{\pi G} \left (\frac{-2\xi \eta}{{(\xi - \eta)}^{3}}\right ),
\label{Bdensity}
\end{equation}
which, in the usual $(r, \theta, \phi)$ polar coordinates, takes the form
\begin{equation}
\rho_{{\rm b}}(r, \theta) =\left(\frac{B_{+} - B_{-}}{4\pi G}\right) \frac{{\sin}^{2}\theta}{r}.
\label{Bpolar}
\end{equation}
\ni The density $\rho_{{\rm b}}$ is positive for $B_{+} > B_{-}$ and zero on the
$z$-axis, as is expected of a homogeneous solution of
Eq.~(\ref{diffeqn}).
\begin{eqnarray}
\rho_{{\rm c}}(r, \theta) &= &\left (\frac{C_{-} - C_{+}}{8\pi G}\right )\frac{1}{r^3} - \frac{\delta(r)}{16\pi Gr^2}[\,C_{+}\ln(1+\cos\theta) \nonumber \\ & & - C_{-}\ln(1-\cos\theta)\,],
\label{Cpolar}
\end{eqnarray}
\begin{equation}
\rho_{{\rm d}}(r, \theta) = \left(\frac{D_{-} - D_{+}}{4\pi
G}\right)\frac{\cos\theta}{r^2}.
\label{Dpolar}
\end{equation}
\ni The density $\rho_{{\rm d}}$ changes sign across the $x$-$y$ plane, unless
we require that $D_{+}$ or $D_{-}$ be discontinuous. All terms in 
$\rho_{{\rm bcd}}$ are unphysical and deserve to be ignored.

\label{lastpage}
\end{document}